\documentclass[twocolumn]{aastex631}

\usepackage{graphicx}
\usepackage{longtable}
\usepackage{amsmath}
\usepackage{booktabs}
\usepackage{multirow}
\usepackage{float}
\usepackage{hyperref}
\usepackage{url}

\received{\today}
\submitjournal{ApJL}

\shorttitle{NIRCam and MIRI Imaging of NGC 7469}
\shortauthors{Bohn et al.}

\graphicspath{{./}{figures/}}

\begin{document}

\title{GOALS-JWST: NIRCam and MIRI Imaging of the Circumnuclear Starburst Ring in NGC 7469}

\email{tbohn002@ucr.edu}

\author[0000-0002-4375-254X]{Thomas Bohn}
\affil{Hiroshima Astrophysical Science Center, Hiroshima University, 1-3-1 Kagamiyama, Higashi-Hiroshima, Hiroshima 739-8526, Japan}

\author[0000-0003-4268-0393]{Hanae Inami}
\affiliation{Hiroshima Astrophysical Science Center, Hiroshima University, 1-3-1 Kagamiyama, Higashi-Hiroshima, Hiroshima 739-8526, Japan}

\author[0000-0003-0699-6083]{Tanio Diaz-Santos}
\affiliation{Institute of Astrophysics, Foundation for Research and Technology-Hellas (FORTH), Heraklion, 70013, Greece}
\affiliation{School of Sciences, European University Cyprus, Diogenes street, Engomi, 1516 Nicosia, Cyprus}

\author[0000-0003-3498-2973]{Lee Armus}
\affiliation{IPAC, California Institute of Technology, 1200 E. California Blvd., Pasadena, CA 91125}

\author[0000-0002-1000-6081]{S. T. Linden}
\affiliation{Department of Astronomy, University of Massachusetts at Amherst, Amherst, MA 01003, USA}

\author[0000-0002-1912-0024]{Vivian U}
\affiliation{Department of Physics and Astronomy, 4129 Frederick Reines Hall, University of California, Irvine, CA 92697, USA}

\author[0000-0001-7291-0087]{Jason Surace}
\affiliation{IPAC, California Institute of Technology, 1200 E. California Blvd., Pasadena, CA 91125}

\author[0000-0003-3917-6460]{Kirsten L. Larson}
\affiliation{AURA for the European Space Agency (ESA), Space Telescope Science Institute, 3700 San Martin Drive, Baltimore, MD 21218, USA}

\author[0000-0003-2638-1334]{Aaron S. Evans}
\affiliation{National Radio Astronomy Observatory, 520 Edgemont Road, Charlottesville, VA 22903, USA}
\affiliation{Department of Astronomy, University of Virginia, 530 McCormick Road, Charlottesville, VA 22903, USA}

\author{Shunshi Hoshioka}
\affiliation{Hiroshima Astrophysical Science Center, Hiroshima University, 1-3-1 Kagamiyama,  Higashi-Hiroshima, Hiroshima 739-8526, Japan}

\author[0000-0001-8490-6632]{Thomas Lai}
\affiliation{IPAC, California Institute of Technology, 1200 E. California Blvd., Pasadena, CA 91125}

\author[0000-0002-3139-3041]{Yiqing Song}
\affiliation{National Radio Astronomy Observatory, 520 Edgemont Road, Charlottesville, VA 22903, USA}
\affiliation{Department of Astronomy, University of Virginia, 530 McCormick Road, Charlottesville, VA 22903, USA}

\author[0000-0002-8204-8619]{Joseph M. Mazzarella}
\affiliation{IPAC, California Institute of Technology, 1200 E. California Blvd., Pasadena, CA 91125}

\author[0000-0003-0057-8892]{Loreto Barcos-Munoz}
\affiliation{National Radio Astronomy Observatory, 520 Edgemont Road, Charlottesville, VA 22903, USA}
\affiliation{Department of Astronomy, University of Virginia, 530 McCormick Road, Charlottesville, VA 22903, USA}

\author[0000-0002-2688-1956]{Vassilis Charmandaris}
\affiliation{Department of Physics, University of Crete, Heraklion, 71003, Greece}
\affiliation{Institute of Astrophysics, Foundation for Research and Technology-Hellas (FORTH), Heraklion, 70013, Greece}
\affiliation{School of Sciences, European University Cyprus, Diogenes street, Engomi, 1516 Nicosia, Cyprus}

\author[0000-0001-6028-8059]{Justin H. Howell}
\affiliation{IPAC, California Institute of Technology, 1200 E. California Blvd., Pasadena, CA 91125}

\author[0000-0001-7421-2944]{Anne M. Medling}
\affiliation{Department of Physics \& Astronomy and Ritter Astrophysical Research Center, University of Toledo, Toledo, OH 43606, USA}
\affiliation{ARC Centre of Excellence for All Sky Astrophysics in 3 Dimensions (ASTRO 3D)}

\author[0000-0003-3474-1125]{George C. Privon}
\affiliation{National Radio Astronomy Observatory, 520 Edgemont Road, Charlottesville, VA 22903, USA}
\affiliation{Department of Astronomy, University of Florida, P.O. Box 112055, Gainesville, FL 32611, USA}

\author[0000-0002-5807-5078]{Jeffrey A. Rich}
\affiliation{The Observatories of the Carnegie Institution for Science, 813 Santa Barbara Street, Pasadena, CA 91101}

\author[0000-0002-2596-8531]{Sabrina Stierwalt}
\affiliation{Occidental College, Physics Department, 1600 Campus Road, Los Angeles, CA 90042}

\author[0000-0002-5828-7660]{Susanne Aalto}
\affiliation{Department of Space, Earth and Environment, Chalmers University of Technology, 412 96 Gothenburg, Sweden}

\author[0000-0002-5666-7782]{Torsten B\"oker}
\affiliation{European Space Agency, Space Telescope Science Institute, Baltimore, MD 21218, USA}

\author[0000-0002-1207-9137]{Michael J. I. Brown}
\affiliation{School of Physics \& Astronomy, Monash University, Clayton, VIC 3800, Australia}

\author[0000-0002-4923-3281]{Kazushi Iwasawa}
\affiliation{Institut de Ci\`encies del Cosmos (ICCUB), Universitat de Barcelona (IEEC-UB), Mart\'i i Franqu\`es, 1, 08028 Barcelona, Spain}
\affiliation{ICREA, Pg. Llu\'is Companys 23, 08010 Barcelona, Spain}

\author[0000-0001-6919-1237]{Matthew A. Malkan}
\affiliation{Department of Physics \& Astronomy, UCLA, Los Angeles, CA 90095-1547}

\author[0000-0001-5434-5942]{Paul P. van der Werf}
\affiliation{Leiden Observatory, Leiden University, PO Box 9513, 2300 RA Leiden, The Netherlands}

\author{Philip Appleton}
\affiliation{IPAC, California Institute of Technology, 1200 E. California Blvd., Pasadena, CA 91125}

\author[0000-0003-4073-3236]{Christopher C. Hayward}
\affiliation{Center for Computational Astrophysics, Flatiron Institute, 162 Fifth Avenue, New York, NY 10010, USA}

\author[0000-0003-2743-8240]{Francisca Kemper}
\affiliation{Institut de Ciencies de l'Espai (ICE, CSIC), Can Magrans, s/n, 08193 Bellaterra, Barcelona, Spain}
\affiliation{ICREA, Pg. Llu\'is Companys 23, 08010 Barcelona, Spain}
\affiliation{Institut d'Estudis Espacials de Catalunya (IEEC), E-08034 Barcelona, Spain}

\author[0000-0002-9402-186X]{David Law}
\affiliation{Space Telescope Science Institute, 3700 San Martin Drive, Baltimore, MD, 21218, USA}

\author{Jason Marshall}
\affiliation{Glendale Community College, 1500 N. Verdugo Rd., Glendale, CA 91208}

\author{Eric J. Murphy}
\affiliation{National Radio Astronomy Observatory, 520 Edgemont Road, Charlottesville, VA 22903, USA}

\author[0000-0002-1233-9998]{David Sanders}
\affiliation{Institute for Astronomy, University of Hawaii, 2680 Woodlawn Drive, Honolulu, HI 96822}

\begin{abstract}

We present \textit{James Webb Space Telescope} (\textit{JWST}) imaging of NGC 7469 with the Near-Infrared Camera (NIRCam) and the Mid-InfraRed Instrument (MIRI). NGC 7469 is a nearby, $z=0.01627$, luminous infrared galaxy (LIRG) that hosts both a Seyfert Type-1.5 nucleus and a circumnuclear starburst ring with a radius of $\sim$0.5 kpc. The new near-infrared (NIR) \textit{JWST} imaging reveals 66 star-forming regions, 37 of which were not detected by \textit{HST} observations. Twenty-eight of the 37 sources have very red NIR colors that indicate obscurations up to A$_{\rm{v}}\sim7$ and a contribution of at least 25$\%$ from hot dust emission to the 4.4$\mu$m band. Their NIR colors are also consistent with young ($<$5 Myr) stellar populations and more than half of them are coincident with the MIR emission peaks. These younger, dusty star-forming regions account for $\sim$6$\%$ and $\sim$17$\%$ of the total 1.5$\mu$m and 4.4$\mu$m luminosity of the starburst ring, respectively. Thanks to \textit{JWST}, we find a significant number of young dusty sources that were previously unseen due to dust extinction. The newly identified 28 young sources are a significant increase compared to the number of \textit{HST}-detected young sources (4---5). This makes the total percentage of the young population rise from $\sim$15$\%$ to 48$\%$. These results illustrate the effectiveness of \textit{JWST} in identifying and characterizing previously hidden star formation in the densest star-forming environments around AGN.

%Although the majority of them also lie at the ends of the previously detected CO molecular gas bar, we find that four of these NIR sources are located outside of it, indicating a region of active star formation that is off-axis from the bar. Overall, the addition of these young regions confirm the age bimodality seen in the star-forming regions of the ring. Moreover, we have increased the number of young sources by a factor of four, raising the total percentage of the young population to $\sim$40$\%$.

%Their near and mid-infrared colors are also consistent with young ($<$5 Myr) stellar populations and significant PAH emission. 
% These newly detected NIR sources are found throughout the ring, and they generally match the locations of strong mid-infrared (MIR) emission sources. 
% Among these newly discovered regions, the 4.4 $\mu$m/1.5 $\mu$m flux ratio is about 70$\%$ higher than the ratio of the previously known regions.

\end{abstract}


\section{Introduction} \label{sec:intro}

NGC 7469 is a nearby ($z=0.01627$) luminous infrared galaxy ($L_{\rm IR,\; 8-1000\mu m}=10^{11.6}L_{\odot}$) that is part of the Great Observatories All-sky LIRGs Survey \citep[GOALS;][]{Armus2009}. NGC 7469 hosts both a Type-1.5 Seyfert nucleus \citep{Landt2008} and a compact ($\rm{r=0.5}$ kpc) circumnuclear starburst ring \citep{Miles1994,Genzel1995,Fathi2015,Song2021}. Extensive multi-band imaging and spectroscopy have been reported in the literature \citep[e.g.,][]{Stierwalt2013,Inami2013,Inami2018}, and its Type-1.5 nucleus is one of the most extensively studied in the sky \citep[e.g.,][]{Behar2017,Linden2019,Larson2020}. Reverberation mapping measurements indicate a black hole mass of $\sim10^7M_{\odot}$ \citep{Peterson2014,Lu2021} and an AGN-driven biconical outflow of highly ionized gas has been observed \citep{Muller2011,Robleto2021,Xu2022}. Radio observations also reveal a nuclear CO molecular bar %(or loosely wound spiral arms)
inside the ring that crosses the AGN \citep{Davies2004,Izumi2015,Izumi2020}.

The circumnuclear starburst ring accounts for about two-thirds of the bolometric luminosity of the galaxy \citep[][]{Genzel1995,Song2021}. \textit{Hubble Space Telescope} (\textit{HST})-based UV-through-NIR spectral energy distribution (SED) fitting of thirty 1.1$\mu$m-selected star-forming regions found in the ring suggests a bi-modality in the stellar population \citep[][hereafter DS07]{DiazSantos2007}. DS07 found that nineteen of the regions are likely of intermediate age (8---20 Myr) and are found in regions of low extinction, $A_V\sim$1.25 mag. Five of the remaining regions were classified as having younger ages (1---3 Myr) and are located in regions of higher extinction, $A_V\sim$3 mag. This younger, obscured population coincides with the peaks in the radio free-free emission \citep[DS07;][]{Orienti2010}. The intermediate-aged population, on the other hand, is responsible for the UV, optical, and NIR continuum emission.

\begin{figure*}
\epsscale{1.0}
\plotone{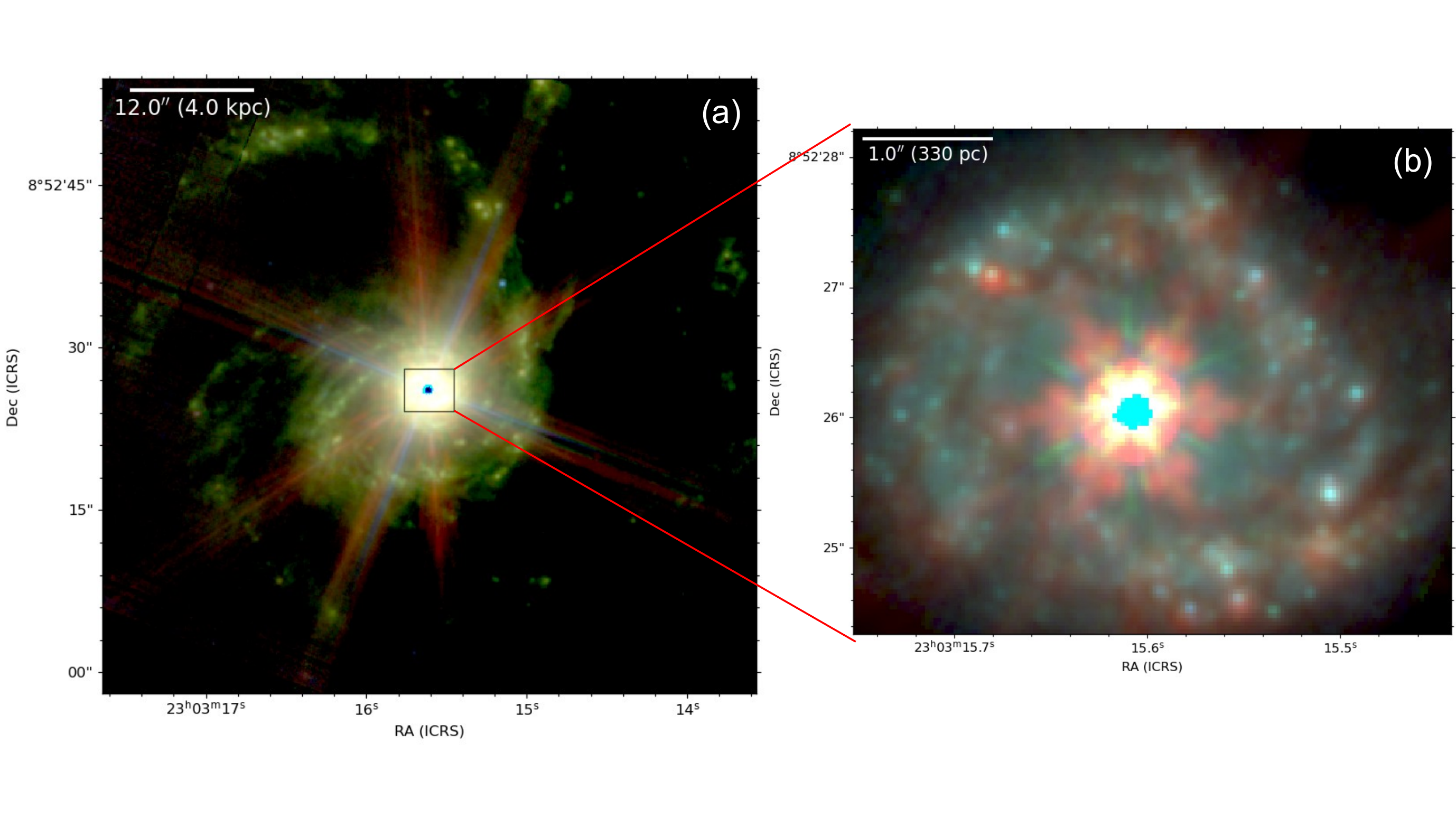}
\caption{False-color images of NGC 7469 taken with MIRI/BRIGHTSKY (\textit{a}) and NIRCam/SUB160P subarray modes (\textit{b}). The MIRI image contains all three observed filters: F560W (blue), F770W (green), and F1500W (red). The central box on the left represents the size of the subarray image on the right. The NIRCam imaging includes the F150W (blue), F200W (green), and F444W (red) filters. Fifty-nine and 22 pixels are saturated in the MIRI/F1500W and NIRCam/F444W filters, which have the most saturation, due to the central AGN, respectively. Note that the AGN is not subtracted in these images. North is up and the East is to the left. \label{fig:RGB_plot}}
\end{figure*}

% For luminous infrared galaxy (LIRGs), star formation triggered by the merger of gas-rich galaxies can account for a large fraction of their enhanced IR luminosity \citep[e.g.,][]{Sanders1996}, and much of this star formation often occurs in compact, central regions \citep{DiazSantos2010,Linden2017,Linden2021}. Multi-wavelength studies have also shown that the central regions of LIRGs are often heavily obscured, necessitating observations at longer wavelengths where the interstellar medium (ISM) is less opaque \citep[e.g.,][]{Inami2010}. Indeed, space-based missions, such as \textit{Spitzer}, \textit{AKARI}, and \textit{Herschel}, have provided a wealth of observations from wide-field imaging to high-resolution spectroscopy \citep[e.g.,][]{Zhao2013,Inami2013,Lu2017}. 

NGC 7469 provides the unique opportunity to study the starburst-AGN connection since it hosts an AGN surrounded by a starburst ring. However, due to the compact nature of this system, it was difficult to achieve both the resolution and sensitivity needed to study the circumnuclear environment in the MIR on sub-kpc scales. With \textit{JWST}, we can now explore the dustiest regions of the starburst ring on these scales.

In this \textit{Letter}, we examine NIRCam and MIRI imaging of the circumnuclear starburst ring in NGC 7469. With the unprecedented spatial resolution and sensitivity of \textit{JWST}, we identify a new set of previously undetected star-forming regions that are heavily obscured by dust. A cosmology of $H_0=70$ km s$^{-1}$ Mpc$^{-1}$, $\Omega_\Lambda=0.72$, and $\Omega_{\rm matter}=0.28$ is adopted. With this cosmology, NGC 7469 is located 70.8 Mpc away, and 1$^{\prime\prime}$ subtends 330 pc.

%Through multi-band photometry, we will measure the flux density and colors of all detected sources, and, based on these measurements, provide an age estimation to a subsample.

\section{Observations} \label{sec:Observations}
%MIRI: F560W, F770W, F1500W
%NIRCAM: F150W, F200W, F335M, F444W

% 2022-06-22 MIRI (#17)
% 2022-07-02 MIRI (#30, subarray)
% 2022-07-01 NIRCAM (#19)
% 2022-07-03 NIRCAM (#27, subarray)

\textit{JWST}/MIRI \citep{Bouchet2015,Rieke2015} imaging was obtained with the F560W (5.6$\mu$m), F770W (7.7$\mu$m), and F1500W (15.0$\mu$m) filters with both the BRIGHTSKY and SUB128 subarray modes. Our analysis was done on the SUB128 subarray data where the full starburst ring was observable in the field of view (FoV: $14.1\arcsec\times14.1\arcsec$). Here, the nucleus remained unsaturated by the AGN. Observations of the SUB128 were taken on 2022-07-02, where the exposure time was 46 seconds for the F560W filter and 48 seconds for the F770W and F1500W filters. Figure \ref{fig:RGB_plot}a shows a false-color image of the MIRI/BRIGHTSKY subarray data.

\textit{JWST}/NIRCam \citep{Greene2016} imaging was taken with the F150W (1.5$\mu$m), F200W (2.0$\mu$m), F335M (3.4$\mu$m), and F444W (4.4$\mu$m) filters. Like MIRI, we use the SUB160P subarray data due to the nucleus being saturated by the AGN in all of the full array images. The entire starburst ring fits within the FoV (short---$5\arcsec\times5\arcsec$, long---$10\arcsec\times10\arcsec$) of the subarray mode and the nucleus was only partially saturated in the F335M and F444W filters. SUB160P observations were taken on 2022-07-03, where the exposure time for each filter was 67 seconds. Figure \ref{fig:RGB_plot}b shows a false-color image of the NIRCam/SUB160P subarray data.

% where the filters are split between short (0.6 - 2.3$\mu$m) and long (2.4 - 5.0$\mu$m) wavelength channels.

% Two subarrays were used: the full array (short: 4 $64\arcsec \times 64\arcsec$ detectors separated by a $4-5\arcsec$ gap, long: $129\arcsec \times 129\arcsec$) and the SUB160P subarray (short: $5\arcsec \times 5\arcsec$, long: $10\arcsec \times 10\arcsec$). Observations were taken on 2022-07-01 (FULL) and 2022-07-03 (SUB160P), where total integration times for all four filters were 0.90 hours and 0.43 hours, respectively.

% and the BRIGHTSKY ($74\arcsec \times 113\arcsec$) and SUB128 ($14.1\arcsec \times 14.1\arcsec$) arrays were used. Observations were taken on 2022-06-22 (BRIGHTSKY) and 2022-07-02 (SUB128), where total integration times for all filters were 0.66 hours and 0.54 hours, respectively. 

\begin{figure*}
\centering
\epsscale{0.93}
\plotone{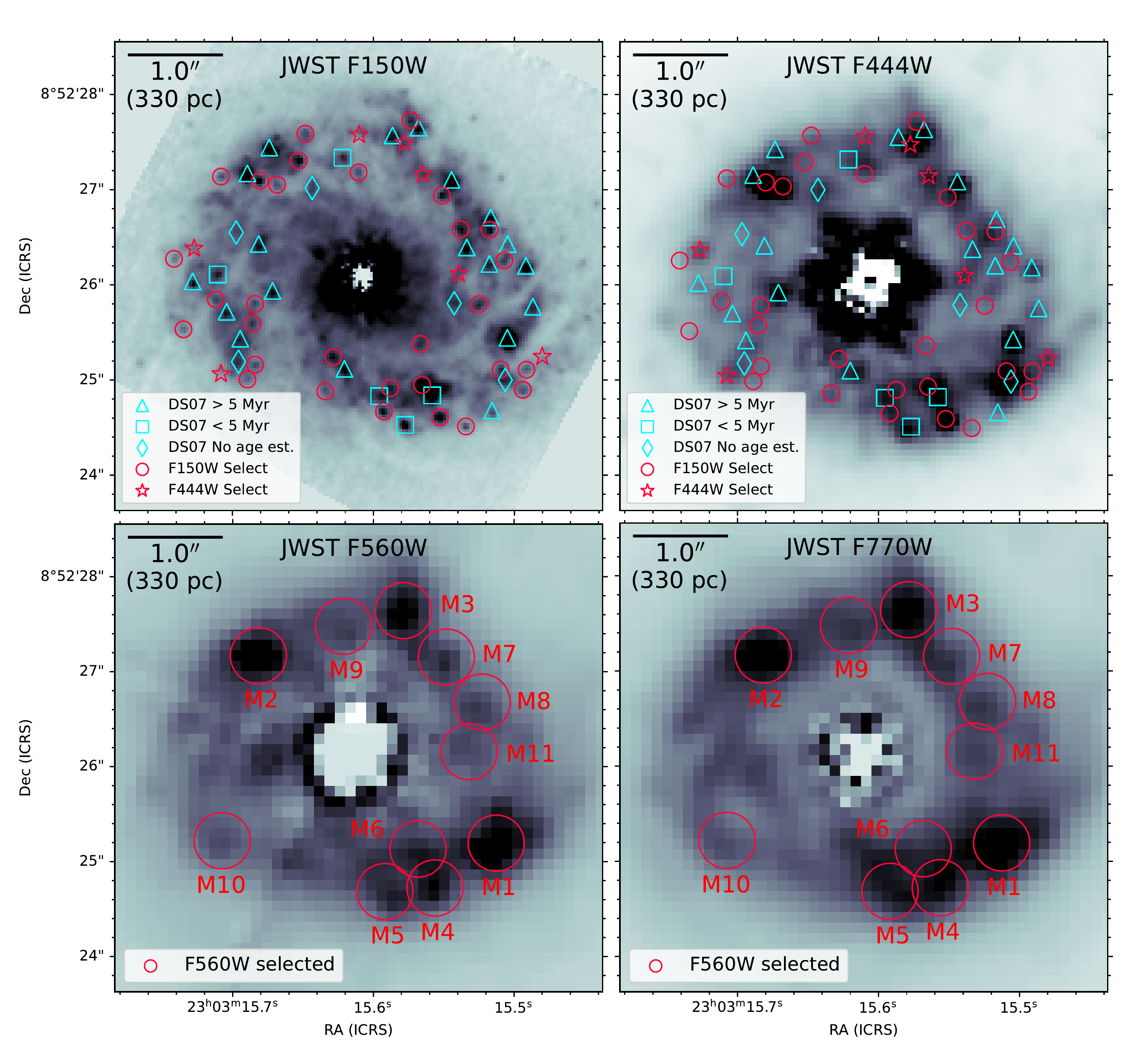}
\caption{NIRCam and MIRI image mosaic of the starburst ring in four filters: NIRCam/F150W, NIRCam/F444W, MIRI/F560W, and MIRI/F770W. The AGN has been masked out in each filter. The source locations for both the NIRCam and MIRI region catalogs are over-plotted. Note that the marker sizes are not representative of the extraction sizes. In the top panels, sources identified with \textit{HST} at 1.1$\mu$m by DS07 that span a range of ages $<$5 Myr (cyan squares) to $>$5 Myr (cyan triangles) are shown. We also include five sources from DS07 where age estimation could not be determined (cyan diamonds). Star-forming regions newly identified with \textit{JWST} are shown as red circles (F150W-selected) and stars (F444W-selected). In the bottom panels, star-forming regions identified in the F560W filter are shown as red circles and are labeled from the brightest, M1, to the faintest, M11. In total, we detect 66 star-forming regions in NIRCam. This more than doubles the 30 sources previously identified by \textit{HST}. \label{fig:mosaic}}
\end{figure*}

All images were processed through the standard \textit{JWST} reduction pipeline \citep{Gordon2015}, and we selected stage 3 calibrated images for analysis. We made corrections to the WCS coordinates to align all the NIRCam and MIRI calibrated images to the Gaia reference frame using publicly available routines\footnote{\url{https://github.com/dancoe/NIRCam}}$^{,}$\footnote{\url{https://github.com/STScI-MIRI/Imaging_ExampleNB}}. Due to the contemporary state of \textit{JWST}, the reduction pipeline is continuously receiving updates. As such, we have applied the most current up-to-date zero-point corrections (CRDS0989) to the flux densities in each NIRCam filter \citep{Boyer2022}\footnote{\url{https://www.stsci.edu/contents/news/jwst/2022/an-improved-nircam-flux-calibration-is-now-available}}.

%For a \textit{JWST} spectroscopic analysis of NGC7469, we refer the reader to L. Armus et al. 2022 for a study of the nuclear emission (in prep.), V. U et al. 2022 for an IFU analysis of the AGN outflow (in prep.) and T. Lai et al. 2022 for an examination of the polycyclic aromatic hydrocarbon (PAH) emission (in prep.). 

\section{Analysis} \label{sec:analysis}

Before source identification was performed, the complex point spread function (PSF) was fit and subtracted from the central source to minimize contamination from instrumental scattered light. This was done by first fitting the AGN with a PSF model using \textsc{GALFIT} \citep{Peng2010}, where the PSF was constructed using the \textsc{WebbPSF} simulation tool \citep{Perrin2014}\footnote{\url{https://www.stsci.edu/jwst/science-planning/proposal-planning-toolbox/psf-simulation-tool}}. The best-fit PSF model was then subtracted, and pixels showing excess emission within the AGN were masked.

Source identification over the starburst ring was then performed, first on the NIRCam/F150W image to take advantage of the high spatial resolution (see Figure \ref{fig:mosaic}). The lack of strong emission lines within its wavelength coverage makes it well-suited to sample the continuum emission. To identify sources, the \textsc{DAOFIND} \citep{Stetson1987} algorithm from the \textsc{photutils} detection package \citep{Bradley2020} was used, where a detected source needed to have a full-width half-maximum (FWHM) greater than or equal to the PSF FWHM of the filter and a peak emission 3$\sigma$ above the local background. This results in a detection limit of $\sim$400 MJy/sr.

A total of 59 star-forming regions were detected in the starburst ring in the F150W filter. About 40 (67$\%$) of these objects appear as unresolved sources. Because the measured FWHM of the PSF of the F150W filter, as determined by field stars, corresponds to $\sim$22 pc, the sizes of the unresolved sources are consistent with those of stellar clusters in nearby galaxies, making them viable candidates for compact clusters \citep{Linden2017,Norris2019,Gillen2021}. We do not refer to the resolved sources as compact clusters because a cluster refers to a distinct grouping of stars, which is best modeled as a single stellar population, and are likely gravitationally bound \citep{Krumholz2018}. For consistency, we refer to all detected sources as ``regions", whether they are resolved or not. We also note that 29 of the F150W-detected sources were selected via \textit{HST}/F110W (1.1$\mu$m) filter imaging in DS07. An additional source, C26, was detected in DS07 but only in the \textit{HST} 1.1$\mu$m band. This source is not detected in the \textit{JWST} images and we exclude it from the analysis.

\begin{deluxetable*}{ccccccccccc}
\tablecaption{Source Photometry}
\label{tab:fluxes}
\tablehead{\colhead{ID} & \colhead{RA} & \colhead{Dec} & \colhead{Dist. from} & \colhead{F150W} & \colhead{F200W} & \colhead{F335M} & \colhead{F444W} & \colhead{F560W} & \colhead{F770W} & \colhead{F1500W}\\ \colhead{ } & \multicolumn{2}{c}{Degrees (ICRS)} & \colhead{Center (kpc)} & \colhead{mJy} & \colhead{mJy} & \colhead{mJy} & \colhead{mJy} & \colhead{mJy} & \colhead{mJy} & \colhead{mJy}}
\startdata
C1 & 345.814664 & 8.8737306 & 0.58 & 0.26$\pm$0.02 & 0.27$\pm$0.02 & 0.22$\pm$0.02 & 0.23$\pm$0.02 & --- & --- & ---\\
C2 & 345.8148249 & 8.874195 & 0.47 & 0.14$\pm$0.02 & 0.13$\pm$0.01 & 0.17$\pm$0.03 & 0.21$\pm$0.02 & --- & --- & ---\\
C3 & 345.8153678 & 8.8742908 & 0.54 & 0.13$\pm$0.02 & 0.14$\pm$0.02 & 0.15$\pm$0.02 & 0.13$\pm$0.02 & --- & --- & ---\\
C4 & 345.8148865 & 8.8735722 & 0.49 & 0.14$\pm$0.02 & 0.17$\pm$0.02 & 0.19$\pm$0.04 & 0.16$\pm$0.02 & --- & --- & ---\\
C5 & 345.8146074 & 8.8739447 & 0.60 & 0.13$\pm$0.01 & 0.14$\pm$0.01 & 0.13$\pm$0.01 & 0.12$\pm$0.01 & --- & --- & ---\\
\hline \\[-1.8ex]
N1 & 345.8149157 & 8.8736028 & 0.44 & 0.11$\pm$0.02 & 0.12$\pm$0.03 & 0.18$\pm$0.03 & 0.16$\pm$0.02 & --- & --- & ---\\
N2 & 345.8148628 & 8.8735086 & 0.57 & 0.10$\pm$0.01 & 0.10$\pm$0.01 & 0.18$\pm$0.02 & 0.26$\pm$0.02 & --- & --- & ---\\
N3 & 345.8152803 & 8.8742575 & 0.45 & 0.07$\pm$0.01 & 0.08$\pm$0.02 & 0.08$\pm$0.02 & $<$0.07 & --- & --- & ---\\
N4 & 345.8147162 & 8.8740557 & 0.50 & 0.07$\pm$0.01 & 0.08$\pm$0.02 & 0.13$\pm$0.03 & 0.11$\pm$0.02 & --- & --- & ---\\
N5 & 345.8153949 & 8.8741989 & 0.48 & 0.07$\pm$0.02 & 0.11$\pm$0.02 & 0.27$\pm$0.03 & 0.44$\pm$0.03 & --- & --- & ---\\
... & ... & ... & ... & ... & ... & ... & ... & ... & ...\\
N36 & 345.8145609 & 8.8736856 & 0.71 & $<$0.02 & $<$0.04 & 0.09$\pm$0.01 & 0.10$\pm$0.02 & --- & --- & ---\\
N37 & 345.8155022 & 8.8736348 & 0.59 & $<$0.02 & $<$0.03 & 0.07$\pm$0.02 & 0.09$\pm$0.02 & --- & --- & ---\\
\hline \\[-1.8ex]
P1 & 345.8153549 & 8.8737623 & 0.30 & $<$0.03 & $<$0.02 & 0.09$\pm$0.03 & $<$0.06 & --- & --- & ---\\
P2 & 345.8152995 & 8.8737334 & 0.33 & $<$0.02 & $<$0.03 & 0.09$\pm$0.02 & $<$0.08 & --- & --- & ---\\
\hline \\[-1.8ex]
M1 & 345.8146537 & 8.8736629 & 0.59 & --- & --- & --- & --- & 3.3$\pm$0.7 & 15.4$\pm$2.9 & 33.9$\pm$4.3\\
M2 & 345.8153581 & 8.8742113 & 0.48 & --- & --- & --- & --- & 3.2$\pm$0.6 & 15.3$\pm$1.7 & 47.1$\pm$4.0\\
M3 & 345.8149285 & 8.8743432 & 0.52 & --- & --- & --- & --- & 2.7$\pm$0.6 & 13.6$\pm$2.8 & 26.6$\pm$2.6\\
M4 & 345.8148351 & 8.8735317 & 0.54 & --- & --- & --- & --- & 2.5$\pm$0.7 & 12.3$\pm$2.0 & 32.5$\pm$2.8\\
... & ... & ... & ... & ... & ... & ... & ... & ... & ...\\
M10 & 345.8154655 & 8.8736699 & 0.56 & --- & --- & --- & --- & 1.3$\pm$0.5 & 6.0$\pm$0.8 & 13.6$\pm$1.8\\
M11 & 345.814734 & 8.8739304 & 0.39 & --- & --- & --- & --- & 1.0$\pm$0.4 & 6.0$\pm$1.2 & 6.7$\pm$3.6
\enddata
\tablecomments{The distance for each source is measured from the central AGN. All flux densities are in units of mJy. Source labels are structured as follows: C---regions first identified by DS07, N---new regions identified in the F150W and F444W filters, P---new regions specifically identified in the NIRCam/F335M filter, M---new MIRI/F560W-identified star-forming regions. \textit{HST}-identified sources (C1-C30) are in order of decreasing \textit{HST} 1.1$\mu$m flux as measured and presented in DS07. All other sources are in order of decreasing 1.5$\mu$m (NIRCam) or 5.6$\mu$m (MIRI) flux density. Upper-limits indicate the source flux is below the 3$\sigma$ detection threshold for that filter.\\
\\
(This table is available in its entirety in a machine-readable form.)}
\end{deluxetable*}
\vspace{-9mm}

To investigate potentially missing sources due to extreme dust obscuration, we ran a similar selection scheme on the F444W image. The F444W filter better samples the continuum than the F335M filter, which may be contaminated by the 3.3$\mu$m polycyclic aromatic hydrocarbon (PAH) feature. An additional seven sources were discovered this way, resulting in a total of 66 star-forming regions detected in the F150W and F444W images (see Figure \ref{fig:mosaic}). For sources that are not detected in a particular filter, we denote the extracted flux as an upper limit (see Table \ref{tab:fluxes}). Lastly, we note that two star-forming regions were detected only in the F335M filter. These are likely PAH-dominated sources and we list them in Table \ref{tab:fluxes} together with the F150W and F444W-selected sources.

% We also refer to these new sources as regions since the PSF FWHM of the F444W filter ($\sim$50 pc) is greater than the maximum star formation cluster size of $\sim$30 pc.

Due to the angular resolution differences between NIRCam and MIRI, a separate sample of star-forming regions were identified in the MIR. Thirteen regions were found in the F560W image using the same method as used above. Of these, two regions were rejected due to intersection with the AGN diffraction spikes. The remaining eleven sources serve as our MIRI sample and their locations are shown in Figure \ref{fig:mosaic}.

\begin{figure*}
\centering
\epsscale{0.93}
\plotone{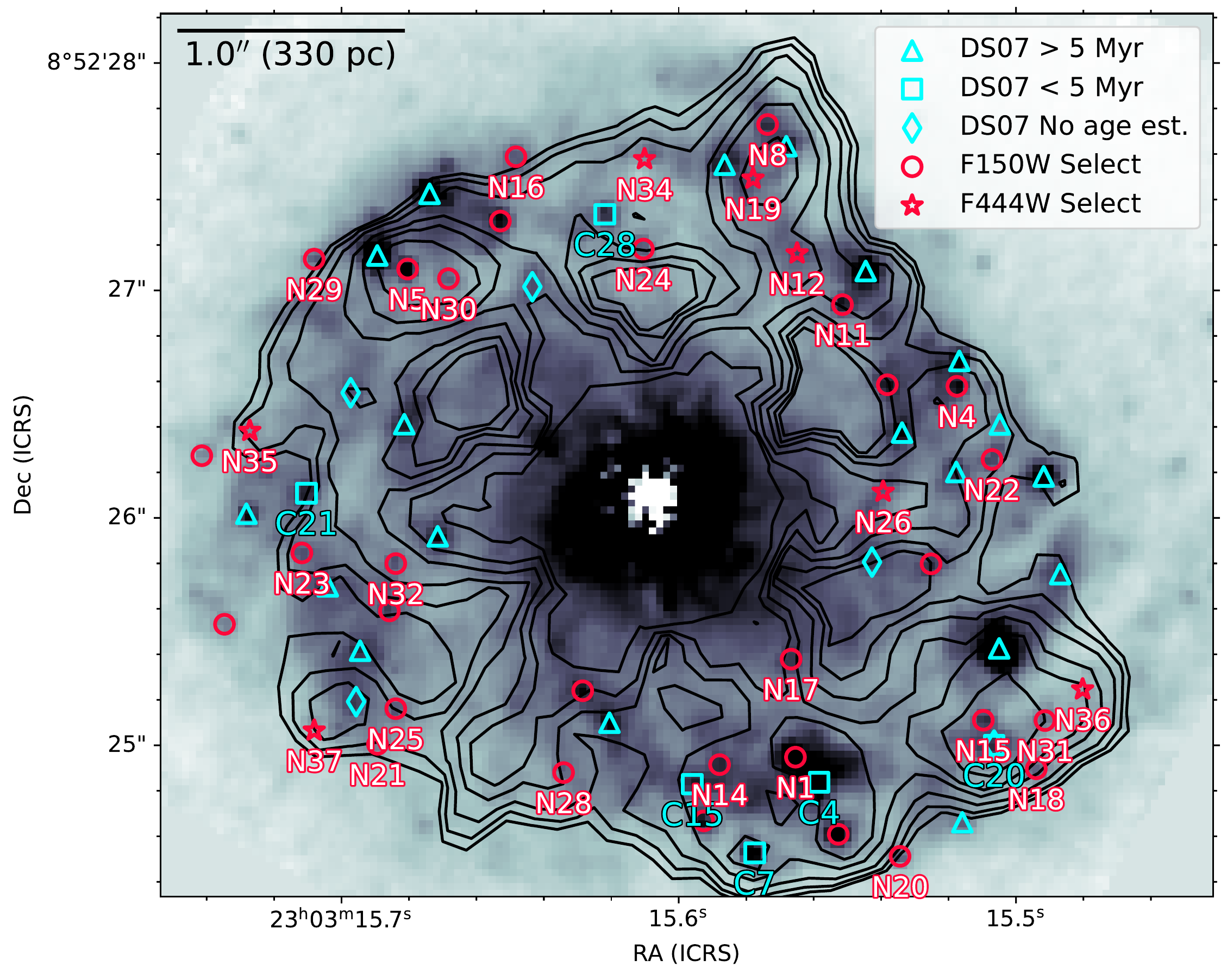}
\caption{NIRCam/F150W image where F150W-selected and NIRCam/F444W-selected star-forming regions with red colors ($\rm{F200W/F150W}>1.1$ or $\rm{F444W/F200W}>2.5$) are labeled. The markers are the same as Figure \ref{fig:mosaic} and the contour levels show the MIRI/F560W emission. Groups of sources with red NIR colors are somewhat clustered in the North and Southwest regions which coincide with the peaks of MIR and radio emission. The markers sizes are not representative of the extraction sizes. \label{fig:contour}}
\end{figure*}

Aperture photometry was performed using the \textsc{photutils} photometry package\footnote{\url{https://photutils.readthedocs.io/en/stable/aperture.html}}. For NIRCam, the radius of the extraction aperture was set to 0.073$^{\prime\prime}$, the largest size at which there is no source overlap in all four filters. For MIRI, the aperture radius for all three filters was set to 0.31$^{\prime\prime}$, where source overlap was also kept to a minimum. To perform local background subtraction, background annuli with varying radii were used. The inner and outer radii ranged from 1.1---2.0 and 1.5---4.0 times the radius of each aperture, respectively. Any source overlapping with the background annuli was masked and the three sigma-clipped median of the annuli measurements was used for background subtraction (typically 30--45$\%$ of the source flux). Following this, aperture corrections were applied using the encircled aperture-to-total energy of each filter.\footnote{\url{https://jwst-docs.stsci.edu/jwst-near-infrared-camera/nircam-performance/nircam-point-spread-functions}}$^{,}$\footnote{\url{https://jwst-docs.stsci.edu/jwst-mid-infrared-instrument/miri-performance/miri-point-spread-functions}} The resulting flux densities are listed in Table \ref{tab:fluxes}.

% The sources are labeled according to their identification method and are in order of decreasing F150W flux density. Note that the labeling of the \textit{HST}-identified sources matches that of DS07 and are in order of decreasing \textit{HST} 1.1$\mu$m flux measured in DS07.

%This is because most SF regions in the F560W are unresolved and using an aperture size equivalent to the F560W filter PSF would leave a large portion of the source emission outside the aperture.

\section{Results} \label{sec:results}

The significant improvement in sensitivity of \textit{JWST} over previous NIR and MIR observations is evident from the number of new sources identified. Thirty-seven star-forming regions were newly discovered with NIRCam, with some single \textit{HST} sources being resolved into two or more sources. Additionally, eleven regions are detected in the MIRI/F560W image. In the following sections, we discuss the spatial locations, colors, and estimates of the ages of this new population of star-forming regions.

\subsection{Locations of Sources} \label{subsec:locations}

Figure \ref{fig:contour} displays the locations of the NIRCam star-forming regions. Overall, the 66 detected sources are evenly distributed azimuthally throughout the ring. However, we find that the newly detected \textit{JWST} sources are somewhat clustered towards the North and Southwest regions, which coincide with the peaks of radio emission at the both ends of the nuclear CO molecular bar (see Section \ref{sec:discussion} for further discussion).

% that could be a result of active positive feedback from the two neighboring sources (C6 and N27). A similar scenario of active positive feedback inducing SF could be occurring in other regions, such as N2, N9, and other redder sources (see Section \ref{subsec:colors}).

\begin{figure*}
\centering
\epsscale{0.83}
\plotone{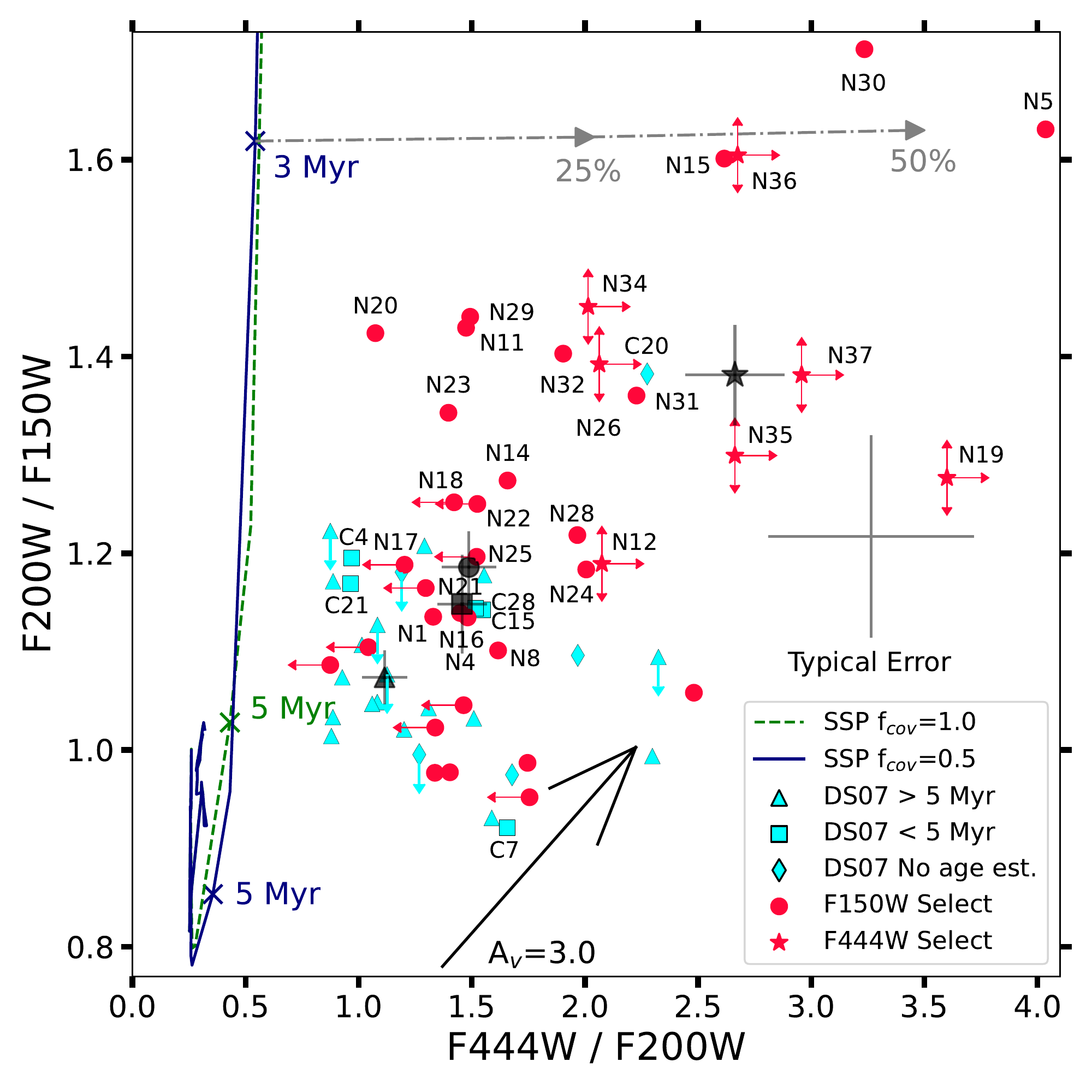}
\caption{NIRCam F200W/F150W vs. F444W/F200W color-color plot of the star-forming regions in the starburst ring of NGC 7469. The markers follow that of Figure \ref{fig:contour}, where red sources, $\rm{F200W/F150W}>1.1$ or $\rm{F444W/F200W}>2.5$, are labeled. Sources with fluxes below the 3$\sigma$ detection threshold are treated as upper-limits for the associated filter. The median colors of each source group are shown with the same markers, but in black. Two single-stellar population models (the green dashed line with f$_{\rm{cov}}$, the fraction of ionizing photons, equal to 1.0 and the dark blue solid line with f$_{\rm{cov}}$=0.5) are also shown, where the crosses denote the locations where the stellar age is 3 Myr and 5 Myr. The typical error and an attenuation vector representing A$_{{V}}=3.0$ are included. The dash-dot line indicates the effect of adding hot dust (800K) to a 3 Myr population. The positions of the arrows indicate fractional contributions of 25$\%$ and 50$\%$ to the F444W flux. The majority of the \textit{JWST} sources with red colors likely have young (3---5 Myr) stellar populations that are highly obscured and contain large amounts of hot dust. \label{fig:Color_plots}}
\end{figure*}

% (\textit{b}) MIRI F1500W/F560W vs. F770W/F560W color-color plot. Colors of nuclei from the GOALS sample using \textit{Spitzer} IRS data are displayed. The data is color-coded based on the 6.2$\mu$m PAH EQW. The position of NGC 7469 as derived from the IRS data is plotted as a black star while measurements of the individual SF regions of this study are shown as empty red circles.

In the MIRI images, seven regions of strong MIR emission (M1---M6, M10) are clearly seen in the ring (see Figure \ref{fig:mosaic}). The locations of these regions match well with previous ground-based MIR observations of the starburst ring \citep{Soifer2003,Almeida2011}. With \textit{JWST}, an additional four new regions are identified in the Northwest (M7, M8, M11) and Northeast (M9). The combined luminosity of these seven regions accounts for $\sim$36$\%$ of the total luminosity of the starburst ring at 5.6$\mu$m.

A comparison of Figures \ref{fig:RGB_plot}b and \ref{fig:contour} reveals that the sources detected by \textit{HST} in DS07 tend to be the bluest and brightest sources in the F150W filter, while \textit{JWST} uncovers the reddest and faintest sources. Indeed, some of the brightest regions in the F444W filter (N5, N15, N19) are not detected by \textit{HST}. The brightest source at 4.4$\mu$m is N5, and is clearly seen as a red source in Figure \ref{fig:RGB_plot}b. Lastly, as shown in Figure \ref{fig:contour}, over half of the newly detected NIRCam sources---the majority of which have redder colors---fall within regions of strong MIR emission.

%One region, m8, is of particular interest. When comparing the F560W to the F770W image, we note a $\sim$90 pc shift to the Southeast in the F770W image. A possible cause of this effect is the presence of an outflowing wind originating from the AGN. IFU studies of NGC 7469 have shown a biconical outflow orientated roughly 90$^{\circ}$ East of North \citep{Muller2011}. Additionally, spectral analysis done by T. Lai et al. (in prep.) find a high [\ion{Ne}{3}]/[\ion{Ne}{2}] ratio, indicative of a strong radiative field. 
% \vspace{8mm}

\subsection{NIR Colors of Sources} \label{subsec:colors}

Figure \ref{fig:Color_plots} shows the F200W/F150W flux ratios against the F444W/F200W flux ratios for all 66 NIRCam sources. The median values for each group of sources, calculated from Bayesian bootstrapping (3,000 iterations and 1$\sigma$ confidence), are shown as the black matching markers. Among the 37 newly identified \textit{JWST} sources, the median F444W/F150W flux ratio is $\sim$40$\%$ higher than that of the 29 sources reported in DS07. The median $\rm{F200W/F150W}$ and $\rm{F444W/F200W}$ flux ratios of the \textit{JWST}-detected sources are also 16$\%$ and 57$\%$ higher than those of the \textit{HST}-detected sources, respectively. In addition, of these 37 new sources, 28 show red colors ($\rm{F200W/F150W}>1.1$ or $\rm{F444W/F200W}>2.5$). However, there are some newly identified regions with blue NIR colors ($\rm{F200W/F150W}<1.1$ and $\rm{F444W/F200W}<2.5$) that are consistent with the colors of the intermediate-aged \textit{HST} sample. These sources were previously missed by \textit{HST}, either due to being below the detection limit or blended with other sources. 

Due to the large range in the resolution of the NIRCam filters used, we cross-convolved each NIRCam image with the PSF of the other filters used in Figure \ref{fig:Color_plots} to bring the image pairs to a common resolution and verify the colors measured in Section \ref{sec:analysis}. The photometric measurements of these convolved images were found to be consistent with those of the non-convolved images, indicating our photometric calibrations are accurate.

%Sources with redder colors, either F200W/F150W $>$ 1.5 or F444W/F200W $>$ 5.0, are labeled. We also label sources with ages estimated to be $<$5 Myr since these will be discussed in Section \ref{subsec:ages}.

\section{Discussion} \label{sec:discussion}

As previously mentioned, 28 of the newly discovered sources show redder NIR colors, $\rm{F200W/F150W}>1.1$ or $\rm{F444W/F200W}>2.5$, than the rest of the sample. To help explain these red colors, we plot in Figure \ref{fig:Color_plots} Yggdrasil single-stellar population (SSP) models\footnote{\url{https://www.astro.uu.se/~ez/yggdrasil/yggdrasil.html}} \citep{Zackrisson2011}, assuming a Kroupa initial mass function \citep{Kroupa2001} and solar metallicity. We include two models, where we altered the fraction of ionizing photons ($\rm{f_{cov}}$) which ionize the cloud from 50$\%$ to 100$\%$. Comparing the colors of our sample to the SSP models, we find that all regions have extinction levels of $A_V\sim$ 1 or greater, which is consistent with the values reported in DS07.
% We also include an attenuation vector as well as two arrows showing different degrees of hot dust contribution to the 4.4$\mu$m continuum at the 3 Myr mark along the f$_\mathrm{cov}$=0.5 SSP model. 

We separate the DS07 sample into three groups, as determined by their SED fits: sources older than 5 Myr, younger than 5 Myr, and those where ages could not be determined. To roughly estimate the ages of the NIRCam star-forming regions, we use their NIR colors, and compare them to those of the \textit{HST}-identified regions with younger stellar populations ($<$5 Myr). The majority of these younger \textit{HST} sources span a specific color space, $1.1<\rm{F200W/F150W}<1.2$ and $0.9<\rm{F444W/F200W}<1.6$, and are redder than most of the other \textit{HST} sources. The median $\rm{F200W/F150W}$ and $\rm{F444W/F200W}$ ratios of these \textit{HST}-detected young sources are 7$\%$ and 35$\%$ higher than the intermediate-aged \textit{HST}-detected sources. One of these young sources, C7, has a significantly lower F200W/F150W value than the rest, but it lacks \textit{HST} photometry longwards of 1.1$\mu$m, which suggests that the inferred age for C7 may be inaccurate.

Most of the newly discovered \textit{JWST} sources are redder in both colors than the young regions catalogued in DS07. Many of these sources are likely very young with significant dust obscuration and hot dust emission affecting their NIR colors. Some regions (N17, N20) fall close to the SSP models and thus likely contain a younger stellar population in a less obscured environment. Other regions (N14, N28) likely host younger stellar populations, but are highly obscured. Their colors can be mainly explained through varying degrees of attenuation, which can be as high as $A_V\sim$ 7. They are also located within regions of strong MIR emission as shown in Figure \ref{fig:contour}. For the reddest sources, $\rm{F444W/F200W}>2.5$, heavy obscuration alone cannot explain the observed colors; a contribution of at least 25$\%$ from hot dust to the 4.4$\mu$m band is needed. The degree to which all these red sources are obscured can provide some further insight into the ages of the stellar populations. Timescales of stellar feedback in young clusters indicate the surrounding gas clouds can be expulsed as fast as 2 Myr \citep{Corbelli2017}, or can take as long as 5 Myr \citep{Messa2021}. A progression of stellar ages has also been linked with obscuration, where younger populations (3 Myr) were shown to be more highly obscured than older populations \citep[5 Myr,][]{Hannon2019}. This would suggest that the reddest sources in the SB ring could be as young or younger than 3 Myr, which is consistent with the ages of dusty SF regions found in this work. Indeed, spectral fitting of MIRI/MRS data reveals that the detected sources are in some of the most obscured locations in the ring \citep{Lai2022}. In addition, star formation rates (SFRs) estimated via the Pf$\alpha$, [\ion{Ne}{2}], and [\ion{Ne}{3}] emission lines indicate that these regions have the highest SFRs in the ring. The 28 newly identified red sources account for $\sim$6$\%$ of the total luminosity of the starburst ring in the 1.5$\mu$m band and $\sim$17$\%$ in the 4.4$\mu$m band.

% We analyze the colors of the MIRI/F560W-selected regions, where many of the NIRCam sources are located, by comparing their MIR colors to those of local LIRGs in the GOALS Survey \citep{Armus2009}. Here, \textit{Spitzer}/IRS spectra of the GOALS sample \citep{Stierwalt2013} were convolved with the MIRI filter transmissions to obtain pseudo-photometry for each filter. All the F560W-selected regions have F1500W/F560W colors between 8---15 and F770W/F560W colors between 5---6, suggesting a very narrow range in PAH equivalent widths (EQWs) which are well within the wider color range (1.5---7.5) shown by the GOALS LIRGs. For the IRS-based photometry of NGC 7469, which includes both the AGN and starburst ring, the AGN contributes significantly to the MIR colors, reducing both the F1500W/F560W and F770W/F560W ratios of the synthetic IRS-based photometry by a factor of 2 compared to the colors of the MIRI regions. 

% As mentioned in Section \ref{subsec:locations}, over half of these red NIRCam sources are located within the MIRI/F560W-selected star-forming regions (see Figure \ref{fig:contour}). The F770W/F560W colors of the seven MIRI-selected sources place them within the range of PAH EQWs that are ascribed to pure star formation \citep{Stierwalt2014}. Indeed, the locations of 2/4 (C4, C15)\footnote{We omit C7 due to its uncertain age estimate.} of the youngest \textit{HST}-selected regions ($<$5 Myr) fall within these regions, further suggesting that these MIR regions, and the NIRCam regions located within, are sites of significant active star formation.

As mentioned in Section \ref{subsec:locations}, radio observations have shown a molecular CO(3-2) gas bar oriented $\sim$45$^{\circ}$ North-of-East that connects the starburst ring through the center \citep{Izumi2015}. Comparison with the MIRI 5.6$\mu$m imaging reveals that the three strongest MIR emission regions (M1---M3) are co-spatial with the radio peaks of the bar \citep[D, B, C of Figure 6 in][]{Izumi2015}. These also coincide with the peak emission seen at 8.4 and 33 GHz \citep{Orienti2010,Song2021}. Analysis of the SEDs of these regions show a relatively steep spectral index ($\sim$0.4---0.9), suggesting a non-thermal origin \citep{Orienti2010}. Since about half of the \textit{HST} and \textit{JWST}-identified sources with younger ages in the Northeast and Southwest regions (N5, N15, N19, etc.) are aligned with these radio peaks, the origin of this non-thermal emission is most likely from supernovae events. If we assume the populations to be 3---5 Myr, then this emission could be from a second wave of OB stars.

Lastly, we note that seven of the red NIRCam sources are located outside of the bar: six in the East/Southeast (N21, N23, N25, N32, N35, N37) and five in the North/Northwest (N4, N11, N12, N22, N26). In the North/Northwest, the NIRCam sources are located in the region of the CO(3-2) emission, which could be an extension of the bar \citep[or a part of a spiral arm;][]{Izumi2015,Izumi2020}. We also detect MIR emission at the same location (M3, M7, M8, M11). In the East/Southeast, weak CO(2-1) and CO(3-2) emission that is co-spatial with the East/Southeastern sources is seen \citep{Davies2004,Izumi2020}. In addition, \citet{U2022} report [\ion{Ar}{2}] strong emission that aligns well with the locations of our MIR sources. As such, these indicate a young stellar population and active star formation are occurring outside the bar.

% Indeed, star formation regions in other LIRGs show a larger spread (1.5 -- 7.5) of F770W/F560W colors (see H. Inami et al. 2022, submitted and A. Evans et al. 2022, submitted), which could indicate that the star formation regions of NGC 7469 are more homogenous than others.

% Based on the locations of these regions being greater than 0.5 kpc from the AGN and having no preferred direction with respect to the AGN, the heating of the dust is likely coming from nearby star formation. In Section \ref{subsec:ages}, we discuss the inferred ages of these red sources.

%This regions, C7, actually has the lowest F200W/F150W ratio of the entire sample.
%Additionally, DS07 do not report measurements with wavelengths longer than 1.1$\mu$m for C7. The other four younger-aged regions, however, do have measurements up to 2.2$\mu$m.

\section{Summary} \label{sec:summary}

In this \textit{Letter}, we present \textit{JWST} NIRCam/SUB160P and MIRI/SUB128 multi-band imaging of the circumnuclear starburst ring of NGC 7469. Our broadband imaging includes NIRCam F150W, F200W, F335M, F444W, and MIRI F560W, F770W, and F1500W filters for a full wavelength coverage of 1.5---15$\mu$m. The significant improvement in sensitivity and resolution at NIR and MIR wavelengths has provided the most detailed look at the ages and colors of the star-forming regions in this starburst ring to date. The main results are:

\textbullet\; We report the detection of 59 star-forming regions in NIRCam/F150W imaging and seven star-forming regions in NIRCam/F444W, for a total of 66 detected regions. This more than doubles the 30 sources previously identified by \textit{HST}. In addition, we detect eleven star-forming regions in the MIRI/F560W imaging.

\textbullet\; Based on NIRCam imaging, the new \textit{JWST}-identified sources tend to be the reddest and faintest sources. Among these newly discovered regions, the median $\rm{F444W/F150W}$ flux ratio is about 40$\%$ higher than that of previously \textit{HST}-identified regions. Their median $\rm{F200W/F150W}$ and $\rm{F444W/F200W}$ flux ratios are also higher by about 16$\%$ and 57$\%$.

\textbullet\; We compare SSP models to the NIR colors of the newly discovered sources to estimate their ages. We find that heavy dust obscuration and hot dust emission are necessary to account for their red colors. Finally, we identify 28 star-forming regions with very red NIR colors that likely have young ($<$5 Myr) stellar populations. These sources account for $\sim$6$\%$ and $\sim$17$\%$ of the total luminosity of the starburst ring at 1.5$\mu$m and 4.4$\mu$m, respectively. The 28 newly-detected, young sources hidden by dust are a significant increase over the 4-5 young sources previously identified by \textit{HST}.

\textbullet\; The MIRI-selected regions contribute $\sim$36$\%$ of the total MIR emission of the starburst ring at 5.6$\mu$m. More than half of these regions lie within the previously known CO molecular bar. Most of the red NIRCam-identified sources also fall within these MIR regions, indicating that with \textit{JWST} we are discovering a large number of heavily obscured sources previously missed by \textit{HST}. 

\textbullet\; The locations of three of the MIRI-selected regions agree well with the 8.4 and 33 GHz emission peaks found in the CO molecular bar. Since the origin of this emission is likely non-thermal, supernovae could be contributing significantly to the radio peaks. We also find four red NIRCam sources in an East/Southeastern CO emitting region that are likely not associated with the bar. These regions could therefore be areas of active star formation that are off-axis from the bar.\\

% \begin{acknowledgments}

We thank the referee for their insightful feedback that helped improve the manuscript. The \textit{JWST} data presented in this paper were obtained from the Mikulski Archive for Space Telescopes (MAST) at the Space Telescope Science Institute. The specific observations analyzed can be accessed at \dataset[https://doi.org/10.17909/1ct2-x706]{\doi{10.17909/1ct2-x706}}. STScI is operated by the Association of Universities for Research in Astronomy, Inc., under NASA contract NAS5–26555. Support to MAST for these data is provided by the NASA Office of Space Science via grant NAG5–7584 and by other grants and contracts. This work is based on observations made with the NASA/ESA/CSA James Webb Space Telescope. These observations are associated with program 1328 and are supported by NASA grant ERS-01328. H.I. and T.B. acknowledge support from JSPS KAKENHI Grant Number JP21H01129 and the Ito Foundation for Promotion of Science. YS is supported by the NSF through grant AST 1816838 and the Grote Reber Fellowship Program administered by the Associated Universities, Inc./ National Radio Astronomy Observatory. The Flatiron Institute is supported by the Simons Foundation. V.U acknowledges funding support from NASA Astrophysics Data Analysis Program (ADAP) grant 80NSSC20K0450. AMM acknowledges support from the National Science Foundation under Grant No. 2009416. S.A. gratefully acknowledges support from an ERC Advanced Grant 789410, from the Swedish Research Council and from the Knut and Alice Wallenberg (KAW) foundation. KI acknowledges support by the Spanish MCIN under grant PID2019-105510GBC33/AEI/10.13039/501100011033. This work was also partly supported by the Spanish program Unidad de Excelencia María de Maeztu CEX2020-001058-M, financed by MCIN/AEI/10.13039/501100011033.

% \end{acknowledgments}

\vspace{5mm}
\facilities{\textit{JWST}/NIRCam, \textit{JWST}/MIRI, \textit{Spitzer}/IRS,
            \textit{HST}/NICMOS}

\software{GALFIT \citep{Peng2010},
          WebbPSF \citep{Perrin2014}}

% \bibliography{Paper.bib}{}
% \bibliographystyle{aasjournal}

\end{document}